\documentclass[%
 reprint,
superscriptaddress,
showpacs,preprintnumbers,
 amsmath,amssymb,
pra,
]{revtex4-1}

\usepackage{graphicx}
\usepackage{dcolumn}
\usepackage{bm}

\newcommand{\be}{\begin{equation}}
\newcommand{\ee}{\end{equation}}
\newcommand{\hata}{\hat{a}}
\newcommand{\beq}{\begin{eqnarray}}
\newcommand{\eeq}{\end{eqnarray}}

\begin{document}


\title{Effects of finite temperature on the robustness of the Mott insulator phase in pseudo one-dimensional Bose-Hubbard Model}

\author{Motoyoshi Inoue}
\email{local-fortokyo@suou.waseda.jp}
\affiliation{Department of Nano-Science and Nano-Engineering, Waseda University, Tokyo 169-8555, Japan}

\author{Keita Kobayashi}
\affiliation{Research Institute for Science and Engineering, Waseda University, Tokyo 169-8555, Japan}

\author{Yusuke Nakamura}
\affiliation{Department of Electronic and Photonic Systems, Waseda University, Tokyo 169-8555, Japan}

\author{Yoshiya Yamanaka}
\affiliation{Department of Electronic and Photonic Systems, Waseda University, Tokyo 169-8555, Japan}





\date{\today}

\begin{abstract}
We study the superfluid-Mott insulator (SF-MI) transition in an one-dimensional optical lattice system, and employ the Bose-Hubbard model in two dimension with a combined potential of an optical lattice in one direction and a confining harmonic trap in the other direction, which we refer to as 
the pseudo one-dimension Bose-Hubbard model. There some excited states with respect to the harmonic trap are considered. The Mott lobes shrink in the $\mu$ and $J$ directions of the $\mu$-$J$ phase diagram. The shrinkage occurs because the interactions involving the excited states become weaker than that between particles in the ground state. The dispersion of the in-site particle increases because the energy spacing between the eigenstates of the Hamiltonian decreases at finite temperature. The excited states significantly affect the robustness of the MI phase at finite temperate.

\end{abstract}


\maketitle


\section{Introduction}\label{intro}

Since the realization of Bose-Einstein condensates in the cold atomic 
systems \cite{BEC}, many studies both from theoretical and experimental sides have been 
done. Various parameters, say the strength of the inter-atomic 
interaction and the configuration of the confinement potential, are 
experimentally well-controllable in the cold atomic systems \cite{seigyo}, which offer many 
intriguing phenomena such as the superfluid-Mott insulator (SF-MI) transition. 
Moreover, it is possible to realize a low dimensional system simply by 
tightening the confinement potential in one or two directions, 
and it provides a good testing field to study a low dimensional physics.

Experimentally, the SF-MI transition in this system, which 
is the quantum phase transition, has been observed when the depth of the lattice potential is
increased \cite{jikken}, and it was followed by the observations in several systems 
such as the multi-component one \cite{taseibunexam}, the multi-band one \cite{reikiband}, 
and the low-dimensional 
ones \cite{2d-jikken, 1d-3d-jikken}. 
Theoretically, the strongly-correlated bosonic system in optical lattice is well described by 
the Bose-Hubbard model \cite{BHM}, according to which the possibility of the SF-MI transition 
was expected before its experimental observation \cite{teniyosoku}. 
This model has been analyzed in various situations, i.e., at finite temperature 
\cite{finitemp2,finitemp},  for the multi-component systems \cite{multicomp,spin1,spin2}, 
for the multi-band ones \cite{band, band2}, and for the low-dimensional one \cite{1d2d}.

The one-dimensional optical lattice system is realized experimentally in 
a combined potential of a one-dimensional optical lattice  and a cigar-shaped
harmonic trap. When the harmonic trap is tight in the directions perpendicular 
to the optical lattice and the optical lattice itself is deep, 
this system can be described in a good approximation 
by the one-dimensional (1D) Bose-Hubbard model, 
restricted to the first band. While the 1D Bose-Hubbard model without higher bands has been 
investigated in some detail both at zero and finite temperatures \cite{finitemp}, its validity
is not clear at finite temperature.
 It is reported that the contributions of 
the higher excitations of the harmonic trap cannot be neglected for the one-dimensional system with Bose-Einstein condensate
 at finite temperature \cite{Arahata}. 

Our aim of this paper is to investigate the temperature dependence of the SF-MI transition in a one-dimensional optical lattice system.
To this end,
we concentrate on a cold atomic system in two dimension with a combined potential of an optical lattice in $x$-direction and a confining 
harmonic trap in $y$-direction, and suppose that the optical lattice is so deep that the 1D Bose-Hubbard Hamiltonian without higher bands describes the dynamics in $x$-direction, but that some excited states in $y$-direction have to be  taken account of because the harmonic potential is not tight
 enough to forbid transitions into the excited states absolutely. We refer to this model as the pseudo one-dimensional (P1D) Bose-Hubbard model throughout this paper.  The reason for analyzing 
the P1D model here is that while it is simple and easy to analyze, and it can be realized experimentally with a merit that the strength of the harmonic trap is variable. It is also pointed out that the structure of 
the transition terms in the P1D Bose-Hubbard Hamiltonian is similar to those in the spin transition \cite{spin1,spin2} or the one between the Bloch-bands \cite{band,band2}, implying that our model 
serves as a prototype to understand the physics of the low dimensional SF-MI transition in more
 general situations.

We study the P1D Bose-Hubbard model both analytically and numerically in this paper,
and show how the transition terms affect the SF-MI phase diagram at zero and finite temperatures. The main result at zero temperature is the shrinkage of the Mott lobes, which originates from a decrease in the effective on-site interaction due to the transitions between the harmonic states in $y$-direction. One may expect that such deformations of the Mott lobes can be fitted simply by adjusting an effective on-site interaction in the 1D model, but it is not true and they reflect the complex structure of interaction in the P1D model. Although the phase diagram for the P1D model reduces to that for the 1D one in the tight confinement limit, we note that the shrinkage is non-negligible even for comparatively tight confinement. The contributions of the transition in $y$-direction
 is more enhanced at finite temperature. The Mott lobes disappear gradually with an increase in temperature due to the thermal fluctuation. While the disappearance in the 1D model occurs  independently of the value of the chemical potential , it is not so in the P1D model. This is explained as follows: the effective on-site interaction becomes weaker for particles in the excited states,
 and the fraction of the excited particles becomes larger as the chemical potential is increased, which implies that the hopping term becomes relatively significant for the larger
 chemical potential. This way we have the dependence of the chemical potential in the phase diagram, calculated from the P1D model. We have confirmed through our analysis that it is only at very low density and temperature that the 1D Bose-Hubbard model without higher bands and the P1D one give almost the same results, and therefore that the validity of the 1D model is restrictive.

This paper is organized as follows. In Sec.~\ref{kaiseki}, we introduce the P1D Bose-Hubbard model. The phase diagram in the zero hopping limit 
is obtained analytically. In Sec.~\ref{Result}, we show the phase diagram for non-zero 
hopping both at zero and finite temperatures by employing the mean-field approximation.
 Section~\ref{summary} is devoted to summary and 
discussion.

\section{Pseudo one-dimensional Bose-Hubbard model}\label{kaiseki}

We start with the following Hamiltonian to describe the cold Bose atomic gas system
in two dimension with a combined potential of an optical lattice in $x$-direction and a confining 
harmonic trap in $y$-direction as follows:
\begin{eqnarray}
\hat{H}&=&\int \!\! dx dy \bigg[ \hat{\psi}^\dagger(x,y)\Big( -\frac{\hbar^2 }{2m}\frac{\partial ^2}{\partial x^2}+ V_{\text{opt}}(x)\bigg)\hat{\psi}(x,y)\notag\\
& &+\hat{\psi}^\dagger(x,y)\bigg(-\frac{\hbar^2 }{2m}\frac{\partial^2}{\partial y^2}+V_{\text{har}}(y)-\mu \Big)\hat{\psi}(x,y)\notag\\
& &+\frac{g}{2}\hat{\psi}^\dagger(x,y)\hat{\psi}^\dagger(x,y)\hat{\psi}(x,y)\hat{\psi}(x,y)\bigg] \,,
\end{eqnarray}
where $V_{\text{opt}}(x)$, $V_{\text{har}}(y)=m\omega_y^2\,y^2/2$, $\mu$, and $g$ represent 
the optical lattice potential, the harmonic potential, the chemical potential, 
and the coupling constant, respectively. Note that
no harmonic potential in $x$-direction is 
considered. Although we study only this two-dimensional model 
throughout this paper for simplicity, an extension to three-dimensional systems, e.g. one with a combined potential of an optical lattice in $x$-direction and a confining 
harmonic trap in $yz$-plane, is straightforward.

The field operator $\hat{\psi}(x,y)$ is expanded as 
\be
	\hat{\psi}(x,y)=\sum_\ell\hat{\psi}_\ell(x)u_\ell(y)\,,\label{tenkai1}
\ee
in the complete set of wave functions of harmonic oscillation,
\be
	\left( -\frac{\hbar^2 }{2m}\frac{d^2}{dy^2}+V_{\text{har}}(y)  \right) u_\ell(y) = E_\ell u_\ell(y) \,,
\ee
where $E_\ell=\hbar \omega_y\left( \ell + 1/2 \right)\,$ $(\ell=0,1,2,\cdots)$.
Then we expand the field operator $\hat{\psi}_\ell(x)$ in the complete set of 
the Wannier functions $ w_i(x) $ with  site indices $i$ as 
\be
	\hat{\psi}_\ell(x)=\sum_i \hat{a}_{\ell,i} w_i(x) \,.\label{tenkai2}
\ee
We assume here and hereafter that the optical lattice potential is so deep that 
only the states in the lowest band contribute, but take account of the excited
 states ($\ell \neq 0$) 
of harmonic oscillation in $y$-direction.  
Under the tight-binding approximation we obtain the Hamiltonian, which we refer to as the
P1D Bose-Hubbard Hamiltonian,
\be
	\hat{H}=\hat{H}_{\text{hop}}+\hat{H}_{\text{on-site}}\,,\label{hamimoto}
\ee
with
\begin{align}
	\hat{H}_{\text{hop}}&=-J\sum_{\ell,i}\left[
	\hat{a}_{\ell,i}^\dagger\hat{a}_{\ell,i+1}+\hat{a}_{\ell,i}^\dagger\hat{a}_{\ell,i-1}\right]\,,\\
	\hat{H}_{\text{on-site}}&=\sum_i \hat{h}_i\,,\\
\hat{h}_i &= \sum_{\ell}(E_\ell-\mu)\hat{a}_{\ell,i}^\dagger\hat{a}_{\ell,i}\notag\\
	&\quad +\frac 12 \sum_{\ell_1\ell_2\ell_3\ell_4}U_{\ell_1\ell_2\ell_3\ell_4}
	\hat{a}_{\ell_1,i}^\dagger\hat{a}_{\ell_2,i}^\dagger\hat{a}_{\ell_3,i}\hat{a}_{\ell_4,i}\,,
	\label{J=0Hamiltonian}
\end{align}
where
\begin{align}
	J &= -\!\int\!\! dx \,w_i^\ast(x)\left(-\frac{\hbar^2}{2m}\frac{d^2}{dx^2}+V_{\text{opt}}(x)\right)w_{i+1}(x)\,,\label{myJ} \\
	g_{\ell_1\ell_2\ell_3\ell_4} &= g\!\int\!\! dy \,
		u^\ast_{\ell_1}(y)u^\ast_{\ell_2}(y)u_{\ell_3}(y)u_{\ell_4}(y)\,,\label{myg} \\
	U_{\ell_1\ell_2\ell_3\ell_4} &= g_{\ell_1\ell_2\ell_3\ell_4}\!\int\!\! dx\, 
		w_i^\ast(x)w_i^\ast(x)w_i(x)w_i(x)\,.\label{myU}
\end{align}
Note that the on-site interaction coefficient $U_{\ell_1\ell_2\ell_3\ell_4}$ depends on 
$\ell$ (the quantum number of the harmonic oscillation) while the hopping coefficient $J$ is independent of $\ell$.  
In Table \ref{Uratio}, the ratios of the on-site interaction coefficients to the ground state interaction are shown.  All the on-site interaction coefficients involving the excited states are smaller than the interaction coefficient $U_{0000}$.
\begin{table}
\caption{The ratios of the on-site interaction coefficients $U_{\ell_1\ell_2\ell_3\ell_4}$ to the ground state interaction $U_{0000}$.The coefficients $U_{\ell_1\ell_2\ell_3\ell_4}$ have the symmetric property, for example, $U_{0011}=U_{0101}=U_{1100}$, etc.}\label{Uratio}
\begin{ruledtabular}
\begin {tabular}{ccccc}
&$U_{\ell_1\ell_2\ell_3\ell_4}/U_{0000}$&& &$U_{\ell_1\ell_2\ell_3\ell_4}/U_{0000}$\\
\colrule
$U_{0011}$&$ 1/2$ &&$U_{0002}$& $-\sqrt{2}/4$ \\
 $U_{1111}$& $3/4$ && $U_{0022}$& $3/8$ \\
 $U_{0112}$&$ \sqrt{2}/8$& &$U_{0222}$&$\sqrt{2}/32$ \\
 $U_{1122}$& $7/16$ &&$U_{2222}$& $41/64$
\end{tabular}
\end{ruledtabular}
\end{table}

The harmonic frequency $\omega_y$ representing the strength of the harmonic potential 
comes into the dynamics of our model in two ways, namely through the energy-level spacing
 $\Delta E_y
= E_{\ell+1}-E_{\ell}=\hbar \omega_y$ and the on-site interaction $U_{\ell_1\ell_2\ell_3\ell_4}$. 
The latter dependence
is expressed as  $U_{\ell_1\ell_2\ell_3\ell_4}\sim \sqrt{\omega_{y}} \,$ universally for 
any $\ell$, and is absorbed into rescaling the phase diagram by using the parameters $\mu/U_{0000}$
 and $J/U_{0000}$. Note that it does not mean that the P1D model reduces to the 1D model by such rescaling. We  pay little attention to this trivial $\omega_y$ dependence, but focus on the nontrivial effect of the transitions between the harmonic states, which is absent in the 1D model.

We assume that the confinement harmonic potential is relatively tight but is not so tight as
 all the excited states can be neglected. 
Then, the infinite sum over $\ell$ can be approximated by a finite sum of some lower energy states. Because of the even-odd selection rule for the on-site interaction
 coefficient $U_{\ell_1\ell_2\ell_3\ell_4}$, the transition of one particle from the ground state to the first excited one  is forbidden. 
The transition with the lowest energy from the ground state is either the simultaneous
one of
 two particles to the first exited state or the one of one particle to the second excited state.
Hence, in order to estimate effects of the excited states at a minimum, we have to sum up the states not with $\ell=0,1$ but with $\ell=0,1,2$ at least, and will take this minimal sum
throughout this paper.  

Before full numerical calculations, 
let us turn to the zero hopping limit, $J=0$. 
Then the hopping Hamiltonian $\hat{H}_{\text{hop}}$ in Eq.~(\ref{hamimoto}) is absent,
 and the total Hamiltonian $\Hat{H} =  \hat{H}_{\text{ on-site}}$ is entirely separable with respect
 to the site index $i$.
Note that $\Hat{h}_i$ in $\hat{H}_{\text{on-site}}$ commutates with the in-site particle number operator at the $i$-site, $\hat{n}_i=\sum_\ell\hata_{\ell,i}^\dagger\hata_{\ell,i}\,$, 
and hence the ground state of the total system is obtained simply by the diagonalization 
with a fixed in-site particle number. 
Hereafter, the trivial subscript $i$ is omitted.

Explicitly we set up the following eigenequation for each fixed in-site particle number $n$
and seek the lowest energy eigenstate,
\be
{\hat h} |\Psi\rangle_n =E_n |\Psi\rangle_n \, ,
\ee
where the state $|\Psi\rangle_n$ is expanded as
 \be
 	|\Psi\rangle_n = \sum_{n_0\,n_1\,n_2} \delta_{n, n_0+n_1+n_2}\; g(n_0, n_1, n_2)|n_0, n_1, n_2\rangle\,,
 \ee 
with the direct product of the particle number states $|n_\ell\rangle$ with $\ell=0,1,2$.
For illustration we write down the matrix eigenequations for $n=1$ and $n=2$:
\be
\left[ -\mu+
\begin{pmatrix}
0&0&0\\
0&\Delta E_y&0\\
0&0&2\Delta E_y
\end{pmatrix}
 \right ] \Phi_1 =E_1  \Phi_1  \label{E1} \, ,
\ee
with
\be
\Phi_1 = \begin{pmatrix}
g(1,0,0)\\
g(0,1,0)\\
g(0,0,1)
\end{pmatrix}\,,
\ee
and
\begin{widetext}
\be
\left[-2\mu+
\begin{pmatrix}
U_{0000}&\sqrt{2}U_{0002}&U_{0022}&U_{0011}&0&0\\
\sqrt{2}U_{0002}&2\Delta E_y+2U_{0022}& \sqrt{2}U_{1122}&\sqrt{2}U_{0112}&0&0\\
U_{0002}& \sqrt{2}U_{1122}&4\Delta E_y+U_{2222}& U_{1122}&0&0\\
U_{0011}&\sqrt{2}U_{0112}&U_{1122}&2\Delta E_y+U_{1111}&0&0\\
0&0&0&0&\Delta E_y +U_{0011}& 2U_{0112}\\
0&0&0&0&2U_{0112}& 3\Delta E_y+2U_{1122}
\end{pmatrix}
\right]
\Phi_2=E_2 \Phi_2 \, , \label{E2}
\ee
\end{widetext}
with 
\be
\Phi_2=
\begin{pmatrix}
g(2,0,0)\\
g(1,0,1)\\
g(0,0,2)\\
g(0,2,0)\\
g(1,1,0)\\
g(0,1,1)
\end{pmatrix}\, .
\ee
\begin{figure}[!b]
\includegraphics[width=5cm]{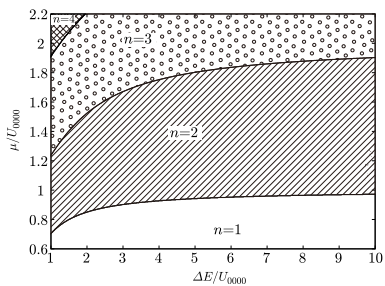}
\caption{The phase diagram in the zero hopping limit, $J=0$, for the P1D Bose-Hubbard model. 
The values of $\mu/U_{0000}$ on the  phase boundary line approach in the tight confinement limit, $\Delta E_y/U_{0000} \to \infty\,$, to the  ones for the 1D Bose-Hubbard model,
which are $\mu/U_{0000} = 1,\,2,\, \cdots$.
}\label{phasediagram_J=0}
\end{figure}
Equation (\ref{E1}) implies that the lowest eigenvalue of $E_1$ is equal to $E_{1,\text{min}}=-\mu$.
One can calculate the the lowest eigenvalue of $E_2$, denoted by $E_{2,\text{min}}$, 
from Eq.~(\ref{E2}).  Similarly, the lowest energy and its eigenstate for each $n$ are obtained.
The results allow us to draw the phase diagram as indicated in Fig.~\ref{phasediagram_J=0}.
There the boundary between $n=1$ and $n=2$ regions, for example, is a line on which 
$E_{1,\text{min}}= E_{2,\text{min}}$.

\begin{figure}
\includegraphics[width=5cm]{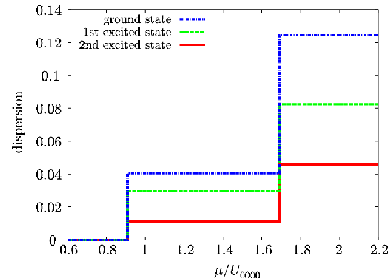}
\caption{(Color online) The dispersion of the particle number in the ground, first excited, and second excited states $\ell=0, 1, 2$ 
for the P1D Bose-Hubbard model with $\Delta E_y/U_{0000}=3.0$.}
\label{J=0n-d}
\end{figure}
\begin{table*}
\caption{The squared coefficient $|g(n_0, n_1, n_2)|^2$ for $\Delta E_y/U_{0000}=3.0$.}\label{joutai}

\begin{ruledtabular}
\begin {tabular}{ccccc}
 in-site particle number&$|g(n, 0, 0)|^2$&$|g(n-1, 0, 1)|^2$&$|g(n-2, 2, 0)|^2$& others \\
\colrule
 $n=2$& $9.84\times 10^{-1}$& $7.88\times 10^{-3}$&$7.47\times 10^{-3}$&$8.37\times 10^{-4}$\\
 $n=3$ &$9.35\times 10^{-1}$&$4.23\times 10^{-2}$&$2.11\times 10^{-2}$&$1.49\times 10^{-3}$\\
 $n=4$& $8.57\times 10^{-1}$& $1.07\times 10^{-1}$&$3.48\times 10^{-2}$&$1.15\times 10^{-3}$\\
\end{tabular}
\end{ruledtabular}
\end{table*}

Obviously, the dispersion of the in-site particle number is equal to zero as a result of
its conservation, and  the state of the whole system is in a Mott phase. But the particle number in each $\ell$-state can fluctuate.   The dispersions of the particle number 
in each $\ell$-state
are shown in Fig.~\ref{J=0n-d}, where they varies stepwise as functions of $\mu$.
For comparison, we recall the corresponding 1D Bose-Hubbard model,
\begin{align}
\hat{H}_{\text{1D}}&=-J\sum_{i}\left[
	\hat{a}_{i}^\dagger\hat{a}_{i+1}+\hat{a}_{i}^\dagger\hat{a}_{i-1}\right]
+ \sum_i {\hat h}_{i,{\text{1D}}}\, , \\
{\hat h}_{i,\text{1D}} &=
-\mu \hat{a}_i^\dagger\hat{a}_i +\frac{1}{2}U_{0000}\,
\hat{a}_i^\dagger\hat{a}_i^\dagger\hat{a}_i\hat{a}_i\,.
\end{align}
In the limit of $J=0$, we consider the eigenequation $\hat{h}_{\text{1D}}|\Psi\rangle_n=E_n|\Psi\rangle_n$ for each fixed $n$, and its lowest eigenvalue is found to
 have a very simple form of
\be
E_n=-n\mu+\frac{n(n-1)}{2}U_{0000} \,,
\ee
which implies that the phase boundaries are at $\mu/U_{0000} = 1,\,2, \cdots\,$ \cite{pethick}. 
We can see from Fig.~\ref{phasediagram_J=0} that 
the existence of the excited states $\ell=1,2$ in the P1D model pushes down the phase boundaries 
towards lower $\mu$. 
It comes from the fact that any on-site interaction involving the excited states $U_{\ell_1\ell_2\ell_3\ell_4}\,$ is smaller than the interaction between ground state particles $U_{0000}$.

In Table~\ref{joutai}, we show the squared coefficients $|g(n_0,n_1,n_2)|^2$ 
for the in-cite particle numbers $n=2,3,4$. 
The probability of the ground state  $|n, 0, 0\rangle$ is the greatest for each $n$, and 
the next leading probabilities, $|n-1, 0, 1\rangle$ and $|n-2, 2, 0\rangle$, 
are nonnegligible. On the other hand, the value in the column ``others", corresponding to a sum 
of contributions of the states with higher energies, is negligibly small 
within our parameter region.  This guarantees the validity of our restriction to the
 energy levels up to $\ell=2$ when the energy-spacing $\Delta E_y/U_{0000}$\, is relatively large. 

\section{Mean-Field Approximation and Numerical Results}\label{Result}

In this section,  we show numerical results of the SF-MI transition and
 the $\mu$-$J$ phase diagram for the P1D Bose-Hubbard model with the hopping Hamiltonian
(the non-vanishing $J$) at zero and finite temperatures. To
deal with the hopping Hamiltonian, we resort to the mean-field approximation\,\cite{finitemp}.

Let us introduce the order parameter 
\be
\Phi_\ell\equiv\langle \hata _\ell \rangle =\frac{\text{Tr}[\hata_\ell\, e^{-\beta \hat{H}}]}{\text{Tr}[e^{-\beta\hat{H}}]}\,,
\ee
where $\beta$ is the inverse temperature $\beta=1/k_{\text{B}}T$
 with the Boltzmann constant $k_{\text{B}}$\,. 
Neglecting the fluctuation terms of the second order, we have an approximate expression of 
\be
\hat{a}_{\ell,i}^\dagger\hat{a}_{\ell,j}\backsimeq 
 \Phi_{\ell,i}^\ast\hata_{\ell,j}+\Phi_{\ell,j} \hata_{\ell,i}^\dagger-\Phi_{\ell,i}^\ast\Phi_{\ell,j}\,.\label{mean-field}
\ee
Then, the approximate hopping Hamiltonian becomes 
\begin{align}
\hat{H}_{\text{hop}}^{\text{(M)}}&=-J\sum_{\ell,i} \Big[(\Phi^\ast_{\ell,i+1}+\Phi^\ast_{\ell,i-1})\hata _{\ell,i}\notag\\
&\quad+(\Phi_{\ell,i+1}+\Phi_{\ell,i-1})\hata^\dagger_{\ell,i}\notag\\
&\quad-\Phi_{\ell,i}^\ast\Phi_{\ell,i+1}-\Phi_{\ell,i}^\ast\Phi_{\ell,i-1} \Big]\,,\label{mean-H}
\end{align}
and the total Hamiltonian under the mean-field approximation is given by $H^{\text{(M)}}=\hat{H}_{\text{hop}}^{\text{(M)}}+\hat{H}_{\text{on-site}}$\,.
We calculate the lowest energy state by diagonalizing the total Hamiltonian, when the order parameter $\Phi_{\ell,i}$ is determined self-consistently
 by minimizing the energy at zero temperature or the free energy 
\begin{eqnarray}
F&=&-\beta^{-1}\ln Z(\beta)\,, \\
Z(\beta)&=&{\text{Tr}}[e^{-\beta\hat{H}^{\text{(M)}}}]\,,
\end{eqnarray}
at finite temperature. 
In this calculation, we consider the excited states up to the second excited state
 as stated in Sec.~\ref{kaiseki}.

\subsection{Results at Zero Temperature }\label{sunb-zero}
\begin{figure}
\includegraphics[width=4.15cm,clip]{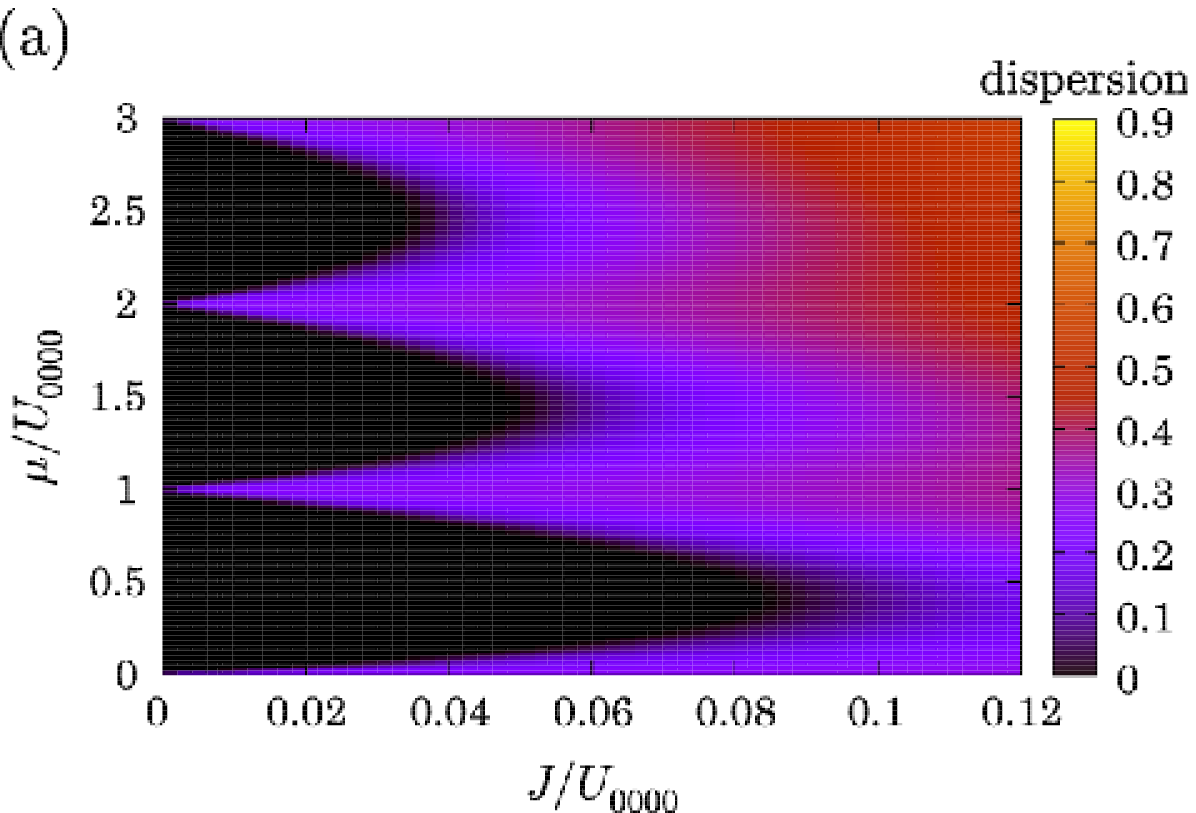}\,\,\,\,\,\,\includegraphics[width=4.15cm]{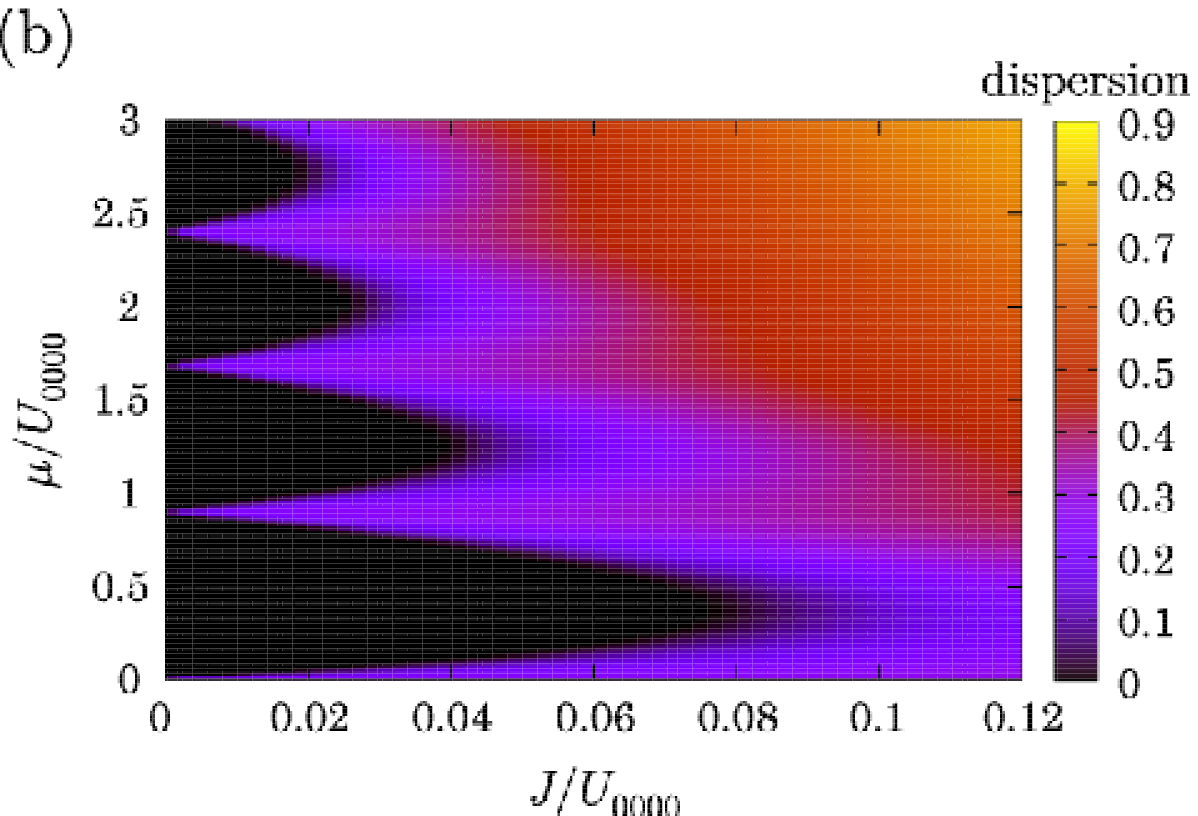}
\includegraphics[width=4.15cm]{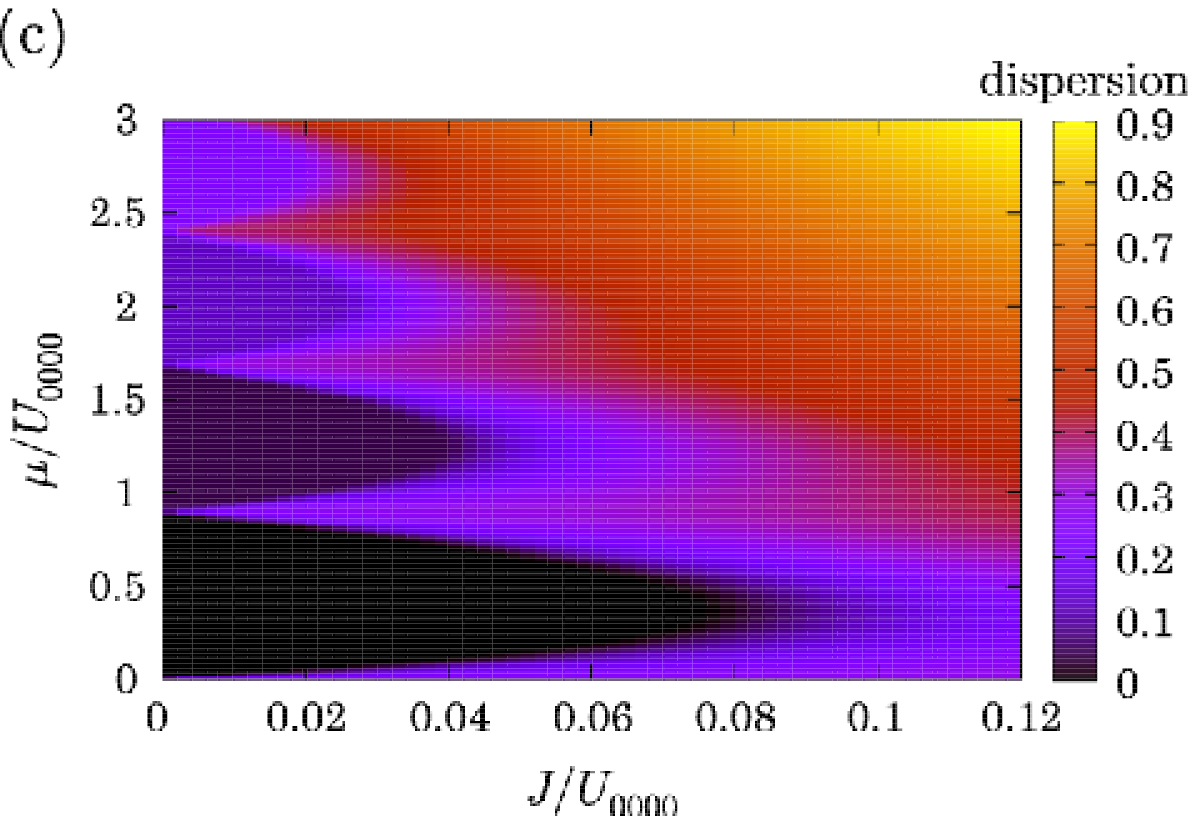}\,\,\,\,\,\,\includegraphics[width=4.2cm]{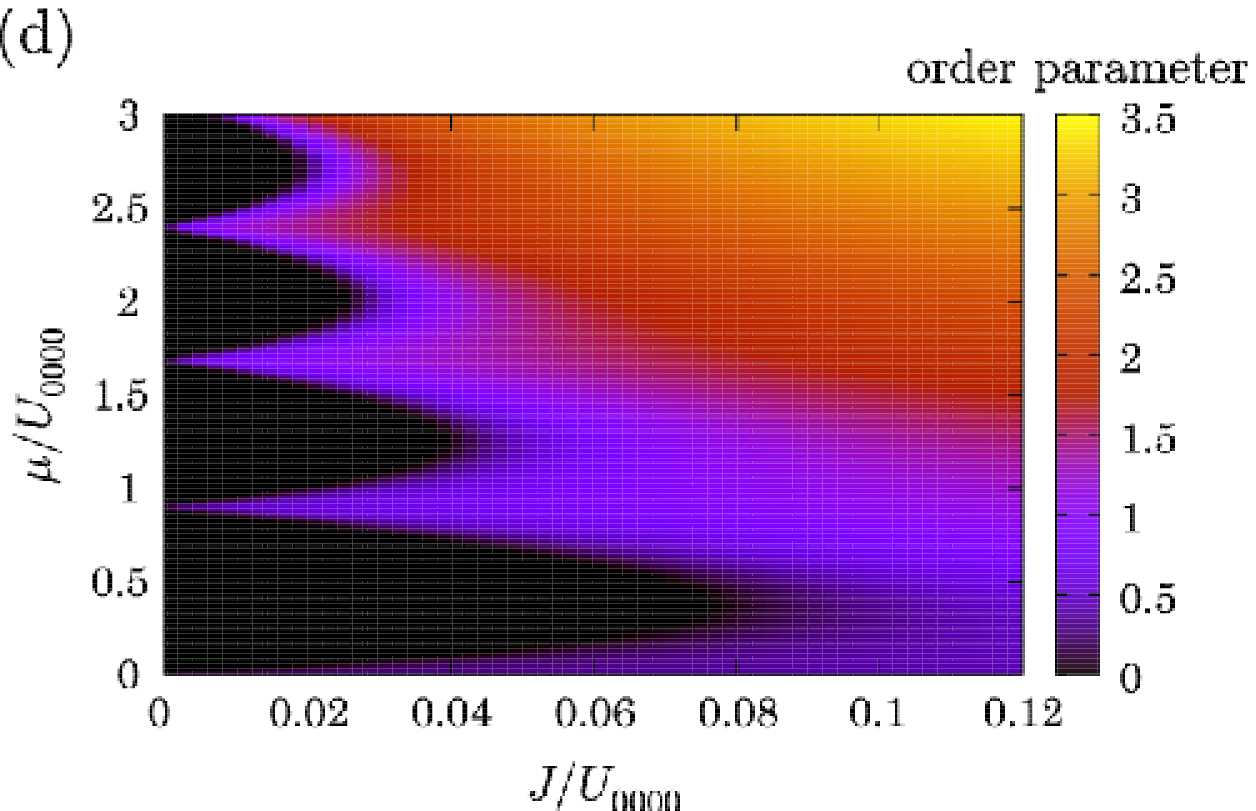}
 \caption{(Color online) (a) The dispersion of the in-site particle number for the 1D model. (b) The dispersion of the in-site particle number. (c) That of the ground state particle number. (d) The absolute square of the ground state order parameter $|\Phi_{0,i}|^2$ for the P1D model with $\Delta E_y/U_{0000}=3.0$ at zero temperature. The black colored regions where the dispersions are zero in (a) and (b) denote the Mott lobes, and the $n$-th lobe from the bottom represents the MI-phase with $n$ in-site particles. The regions where the order parameter vanishes, as depicted in (d), correspond to the Mott lobes in (b).}\label{DE-3.0}

\end{figure}
In Fig.~\ref{DE-3.0}, we show the phase diagrams for the 1D model and for the P1D model with the energy-level spacing $\Delta E_y/U_{0000}=3.0$ at the zero temperature. 
Comparing the result for the P1D model with that for the 1D model in Fig. \ref{DE-3.0}-(a), one can find the following two features.
First, the Mott lobes for the P1D model shrink in the $\mu$ direction in Fig. \ref{DE-3.0}-(b).
The rate of the shrinkage of the Mott lobes for the P1D model increase as the in-site particle number increases.
Secondly, the dispersion of the ground state particle number in the Mott lobes for the P1D model is nonzero and also increases as the in-site particle number increases, as in Fig. \ref{DE-3.0}-(c).
The order parameter does not necessarily become nonzero even in the nonzero dispersion region of the ground state particle number in the Mott lobes, as in Figs. \ref{DE-3.0}-(c) and (d).
The nonzero dispersion of the ground state particle number shows that the particle number of each state is not fixed and can fluctuate, but only the in-site particle number is fixed. 
This is attributed to the decrease of the effective on-site interaction because the on-site interaction coefficients are smaller than the ground state interaction coefficient $U_{0000}$.
The numerical results show that the analysis in Sec.~\ref{kaiseki} can be extended even to the case of $J\neq 0$\,. 

 We see from Figs. \ref{DE-3.0}-(a) and (b) that the Mott lobes for the P1D model also shrink in the $J$ direction. The reason for the shrinkage in the $J$ direction is as follows: 
 the system is favorable to the SF-phase when the ratio of the hopping to the on-site interaction is large.   The effective interaction is smaller than that of the ground state particles, which means that the particles in the excited states can easily be in the SF-phase than the particles in the ground state because then the effect of the hopping is relatively large compared to the on-site interaction. 
As a result, in comparison with the system for the 1D model, the systems for the P1D model is favorable to the SF-phase.

We emphasize that the P1D model is essential in the above considerations since the particles in the excited states play a crucial role.

\subsection{Results at Finite Temperature }\label{finite-temp}

\begin{figure}
\includegraphics[width=4cm]{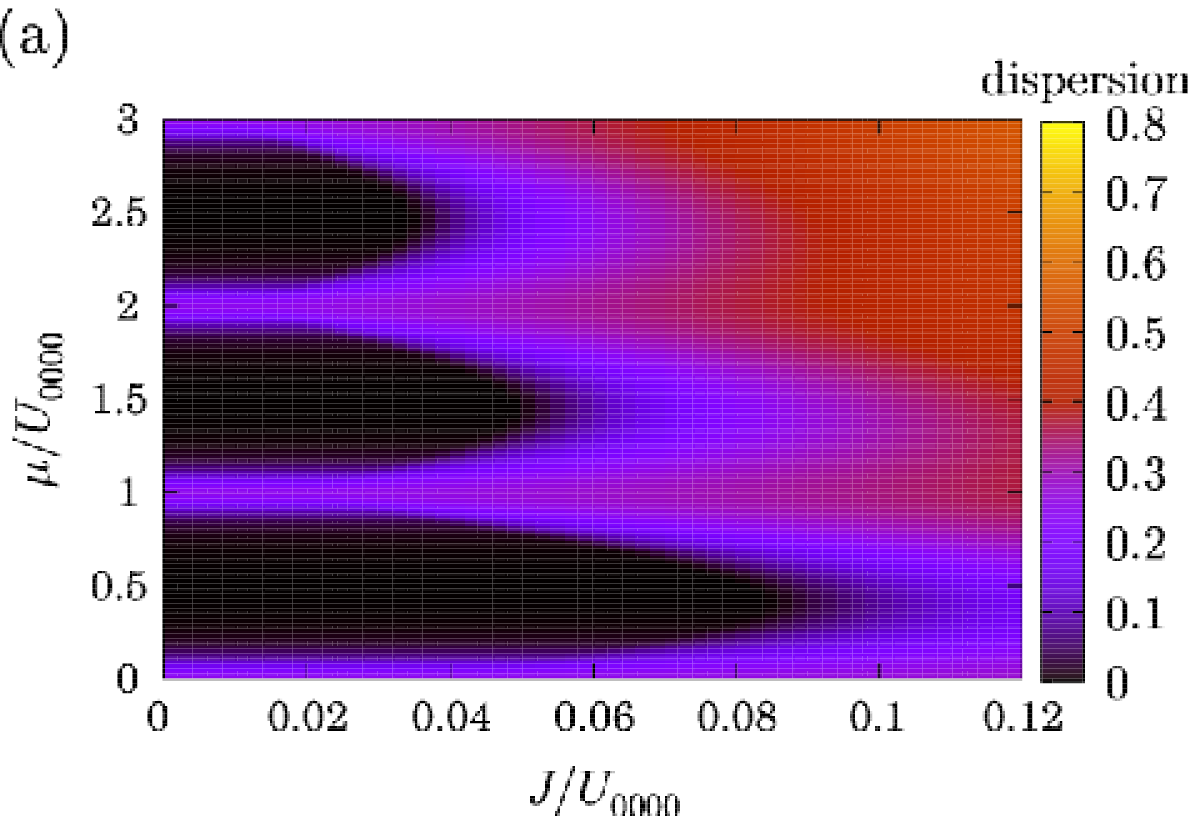}\,\,\,\,\,\,\,\,\,\includegraphics[width=4.15cm,keepaspectratio,clip]{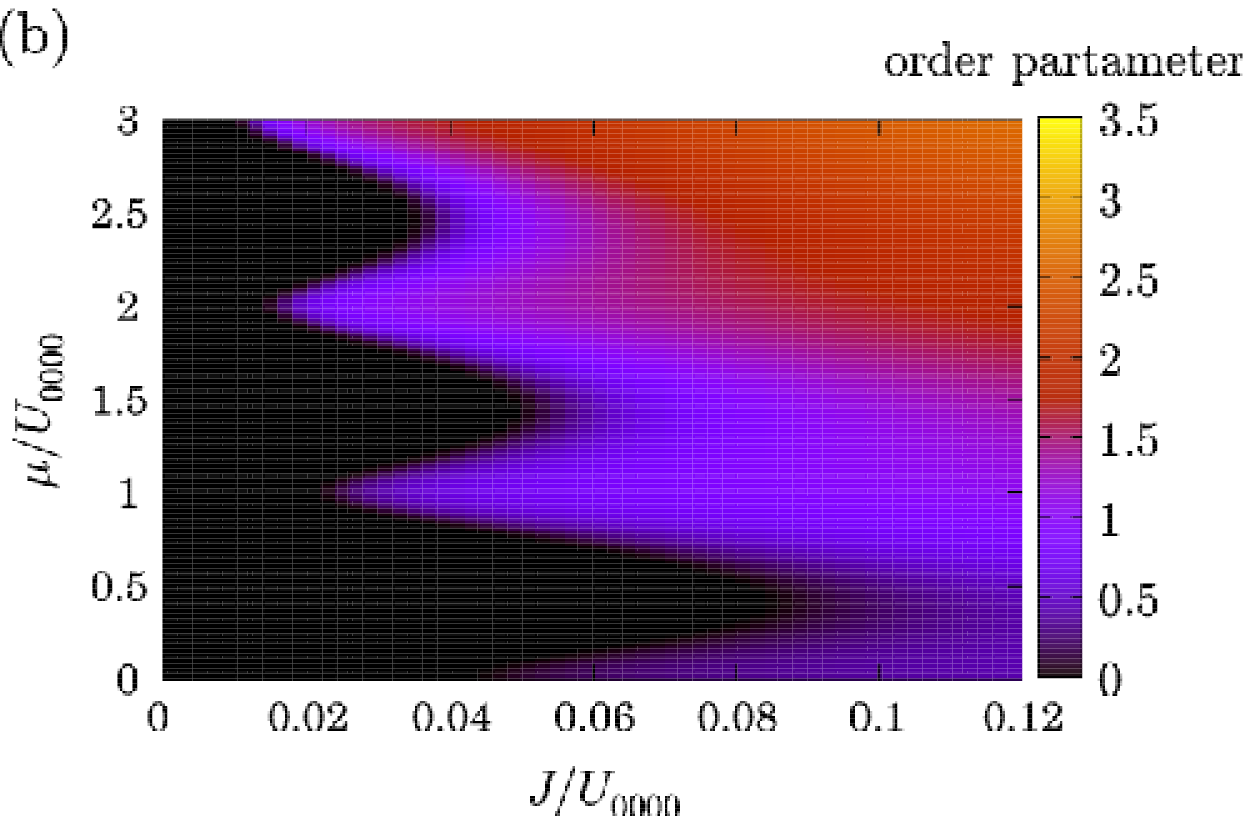}
\includegraphics[width=4cm,keepaspectratio,clip]{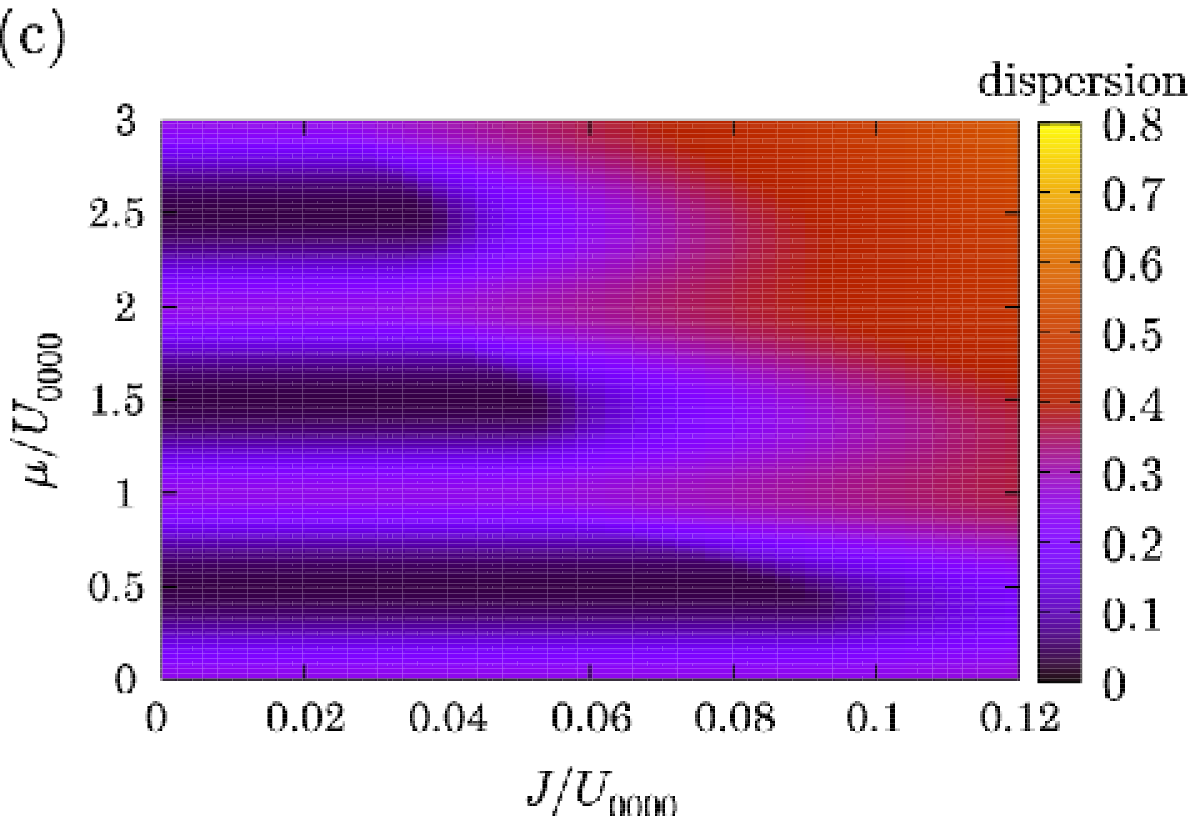}\,\,\,\,\,\,\,\,\,\includegraphics[width=4.15cm,keepaspectratio,clip]{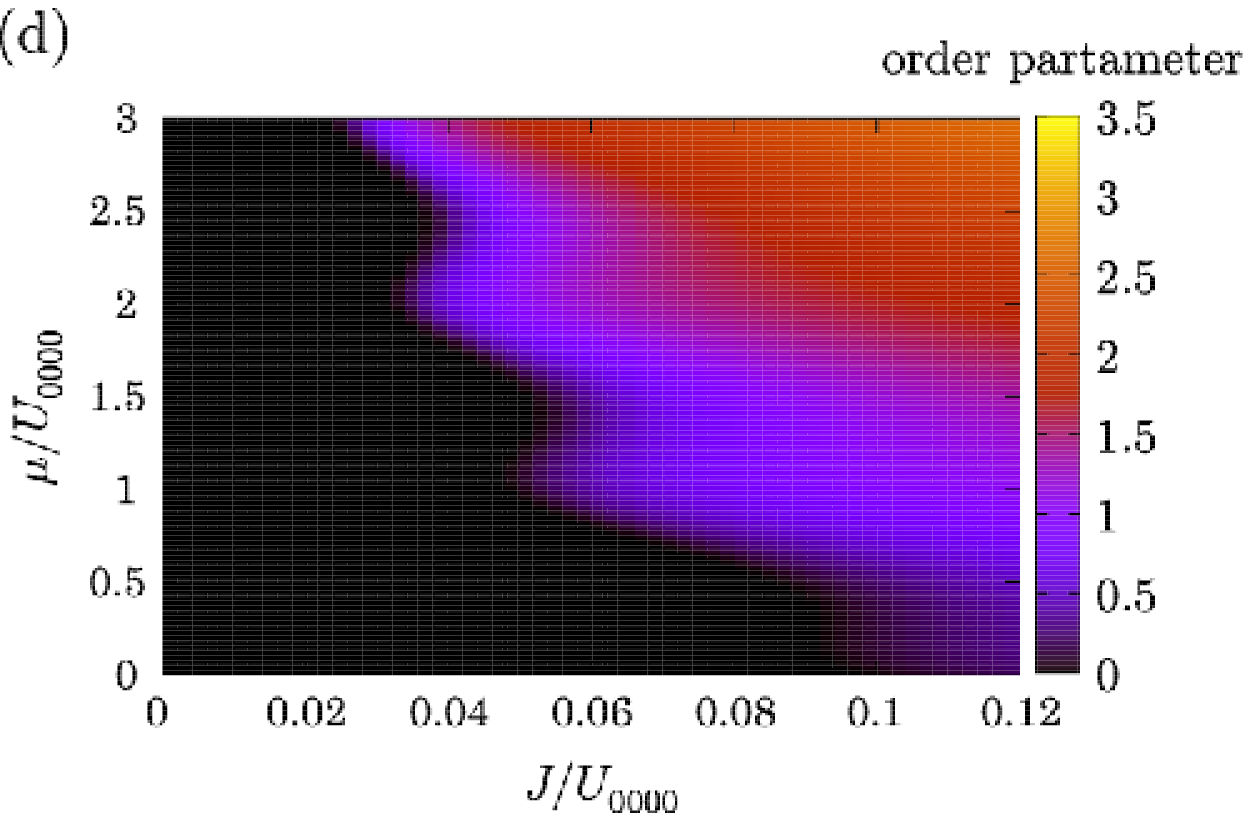}
\caption{(Color online) (a) The dispersion of the in-site particle number and (b) the absolute square of the order parameter with $\beta U_{0000} = 20.0$ for the 1D model. Figures (c) and (d) represent those with $\beta U_{0000} = 8.0\,$, respectively.}\label{1d-8}
\end{figure}

\begin{figure}
\centering\includegraphics[width=4cm]{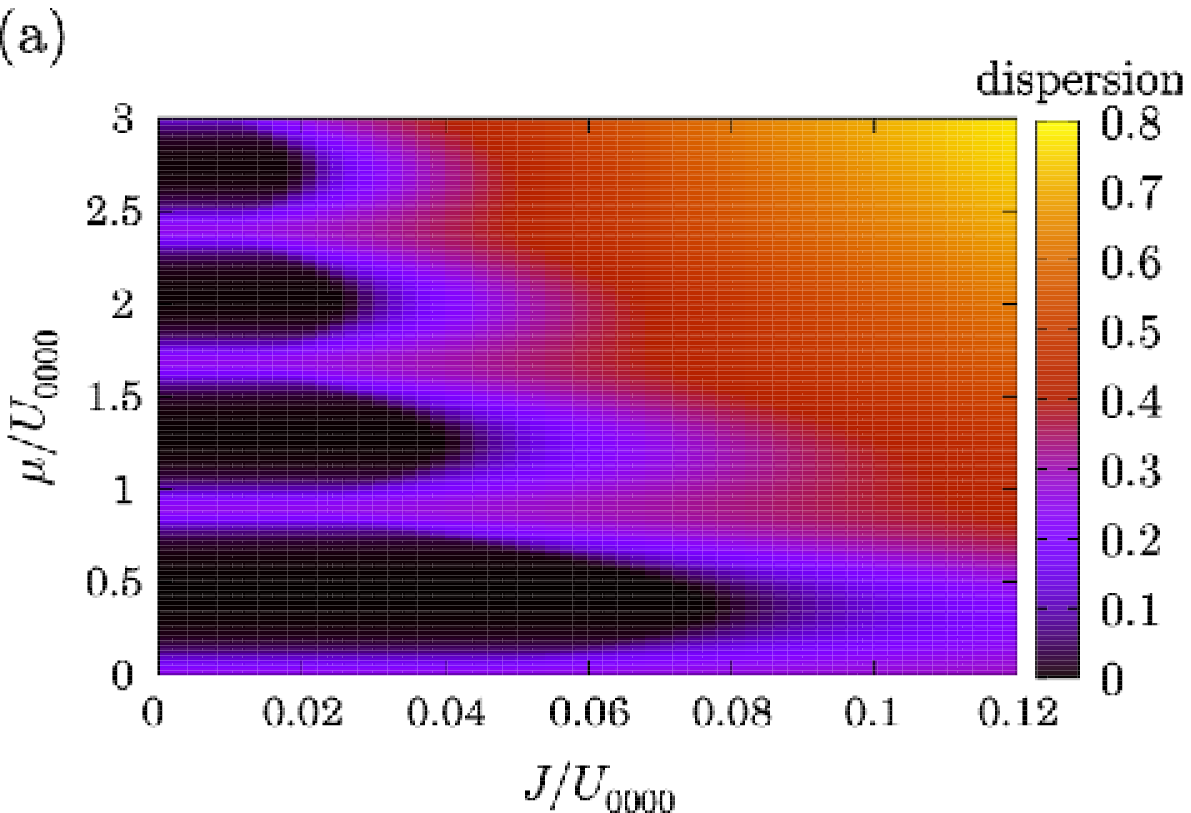}\,\,\,\,\,\,\,\,\,\includegraphics[width=4.15cm]{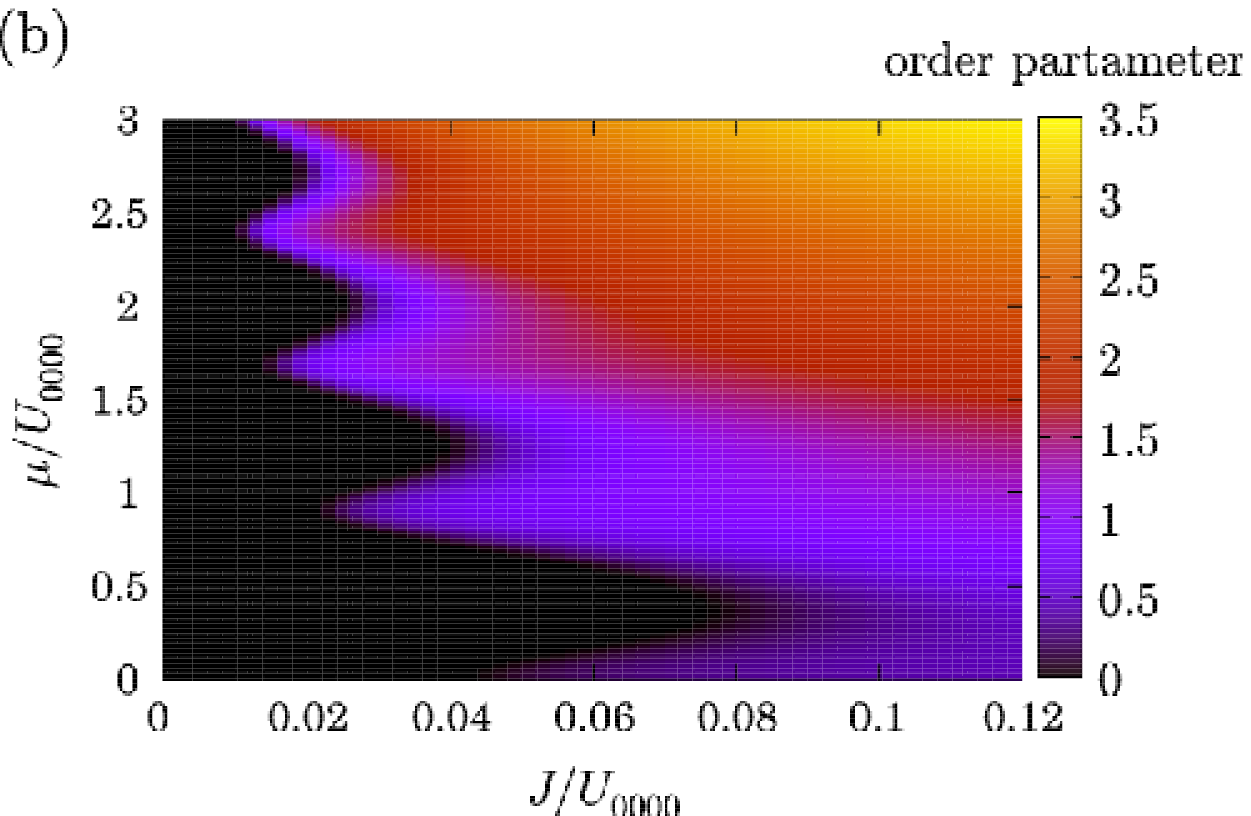}
\includegraphics[width=4cm]{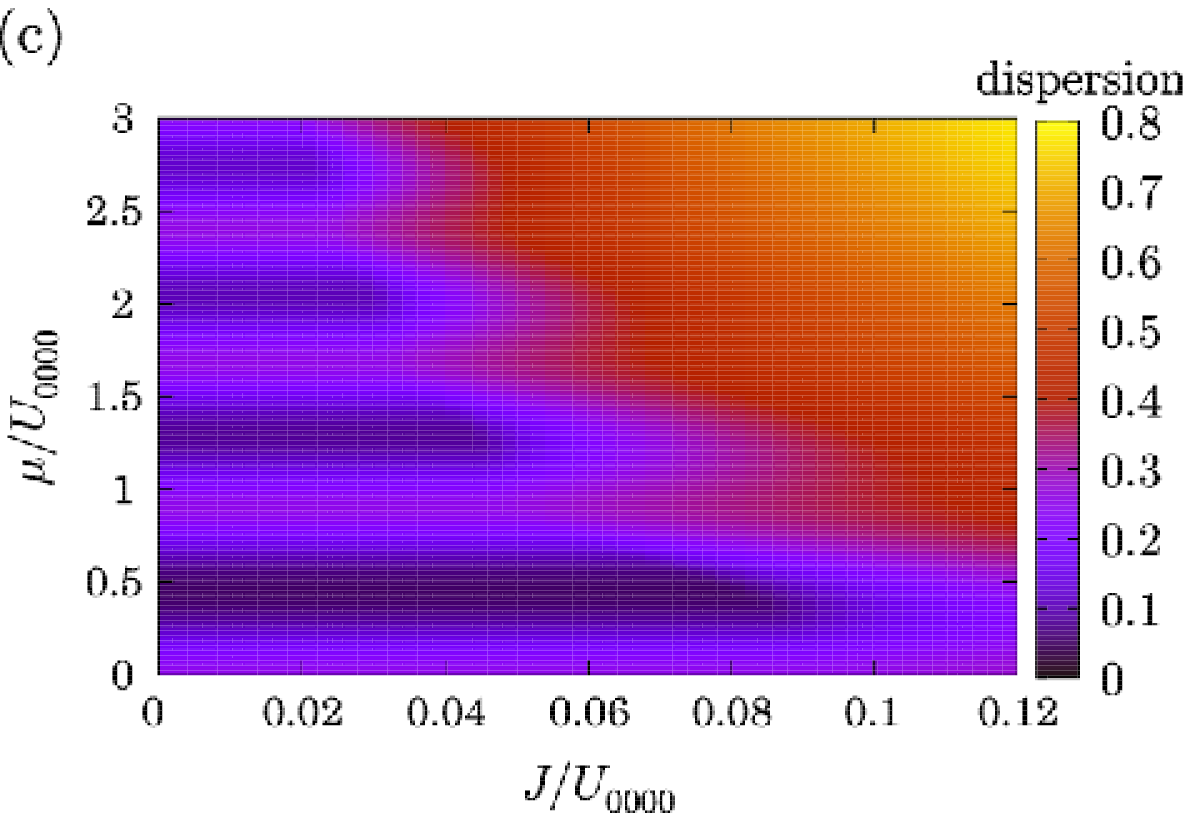}\,\,\,\,\,\,\,\,\,\includegraphics[width=4.15cm]{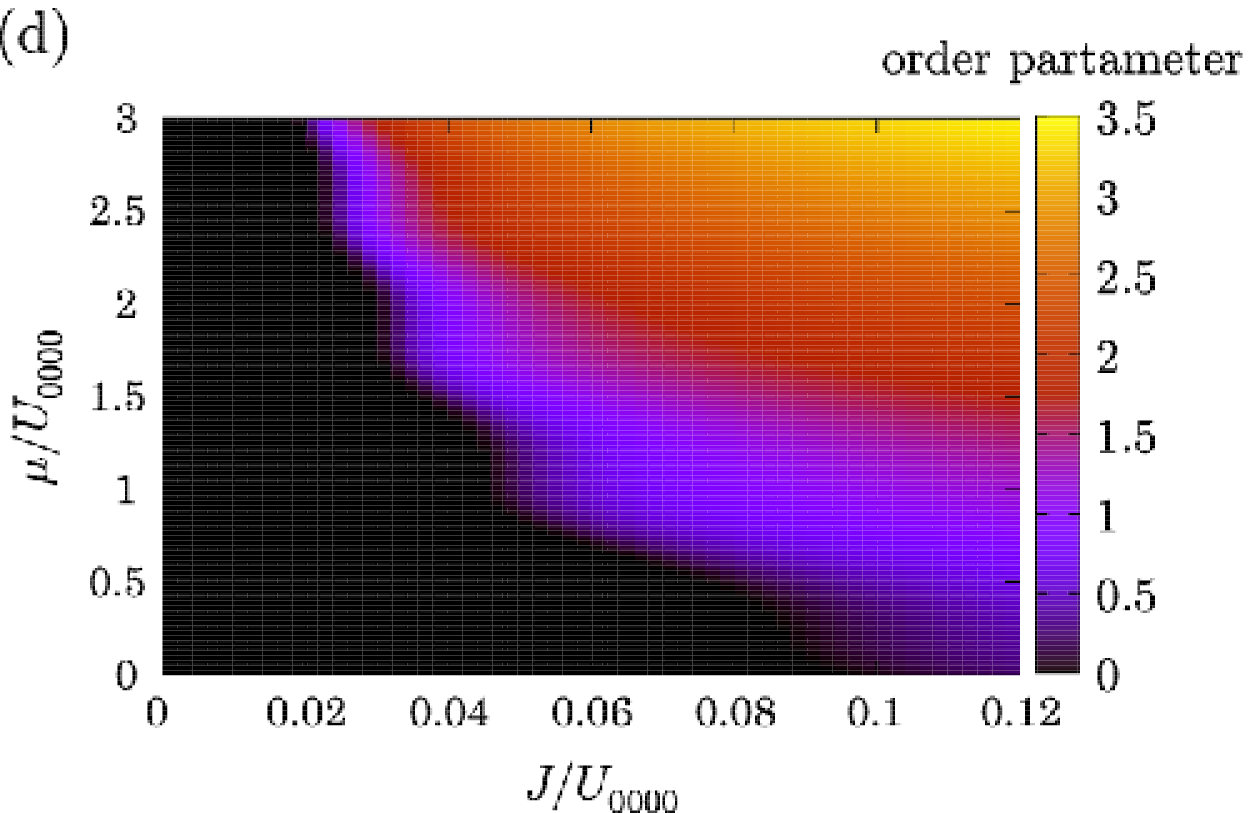}
\caption{(Color online) (a) The dispersion of the in-site particle number and (b) the absolute square of the order parameter with $\beta U_{0000} = 20.0$ for the P1D model. Figures (c) and (d) represent those with $\beta U_{0000} = 8.0\,$, respectively.}\label{2d-8}
\end{figure}

In the previous subsection, we found that the effects of the excited states become more conspicuous
for larger in-site particle numbers. It is therefore anticipated that the excited state
  is more important at finite temperature. 
In this subsection, we analyze the effect of finite temperature for
 the P1D model. 

In Fig.~\ref{1d-8}, we show the particle number dispersions and the order parameters for
 the 1D model at temperatures $\beta \,U_{0000}=20.0$
 and $\beta \,U_{0000}=8.0$. It is seen from Fig.~\ref{1d-8} that the Mott lobes do not fit the region of the vanishing order parameter, the latter is wider than the former. We have some regions where the dispersion is non-vanishing  but the order parameter vanishes, $\Phi_{0}=0$, and interpret that they are in the normal-liquid phase\,\cite{finitemp}.

The phase diagram for the P1D model
 is shown in Fig.~\ref{2d-8}.  There the order parameter is that for the ground state.
As in the 1D model, the vanishing order parameter region covers the vanishing dispersion one.
The normal-liquid region expands as temperature goes up.
In a similar fashion to the zero temperature case, the shrinkage of the Mott lobes for the P1D model also is found at finite temperature. 
The shape of the lobes for the order parameter in Fig. \ref{2d-8}-(b) and (d) is lost at finite temperature more clearly than for the 1D model in Fig.~\ref{1d-8}-(b) and (d).

\begin{figure}
\centering\includegraphics[width=4.4cm]{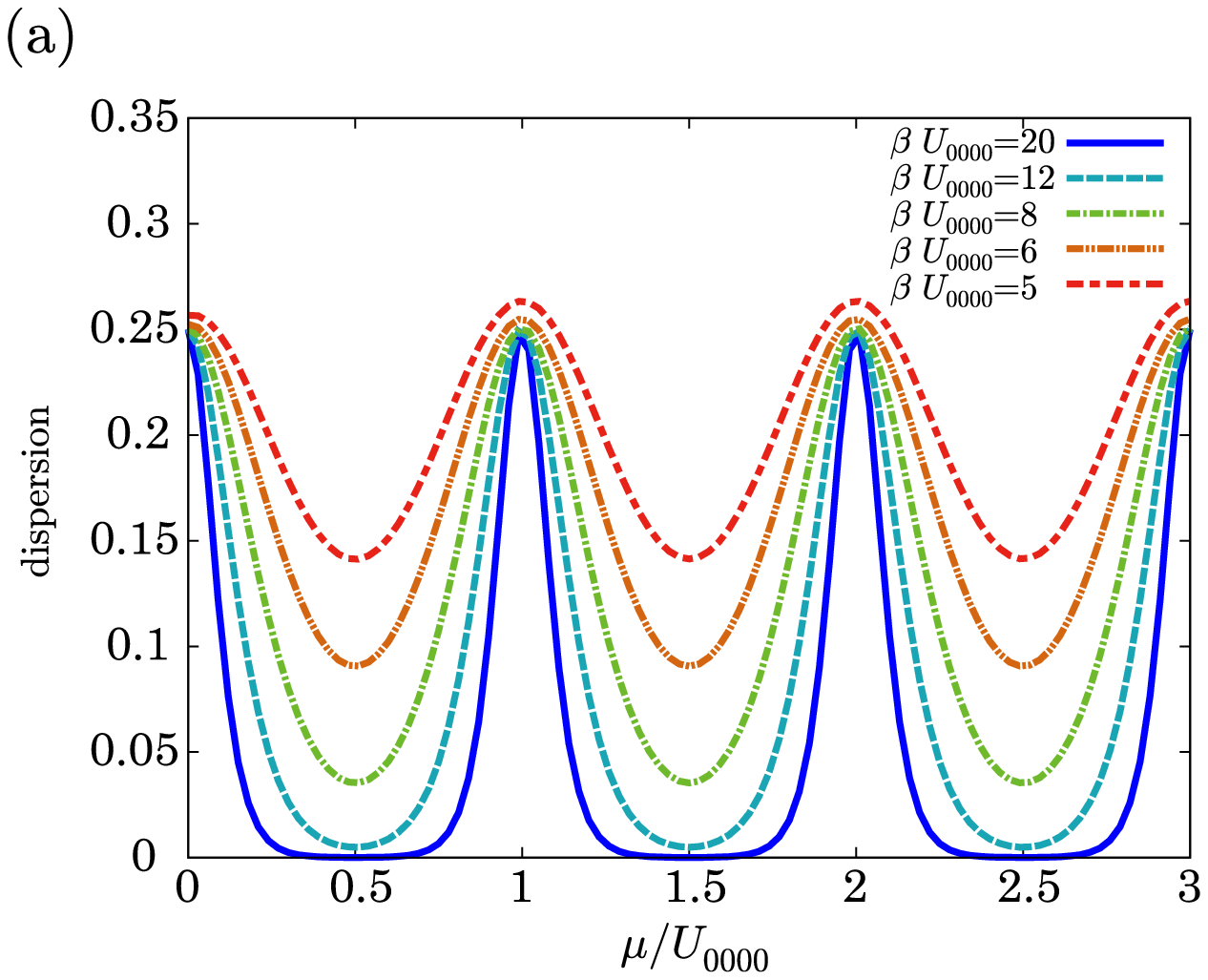}\includegraphics*[width=4.4cm]{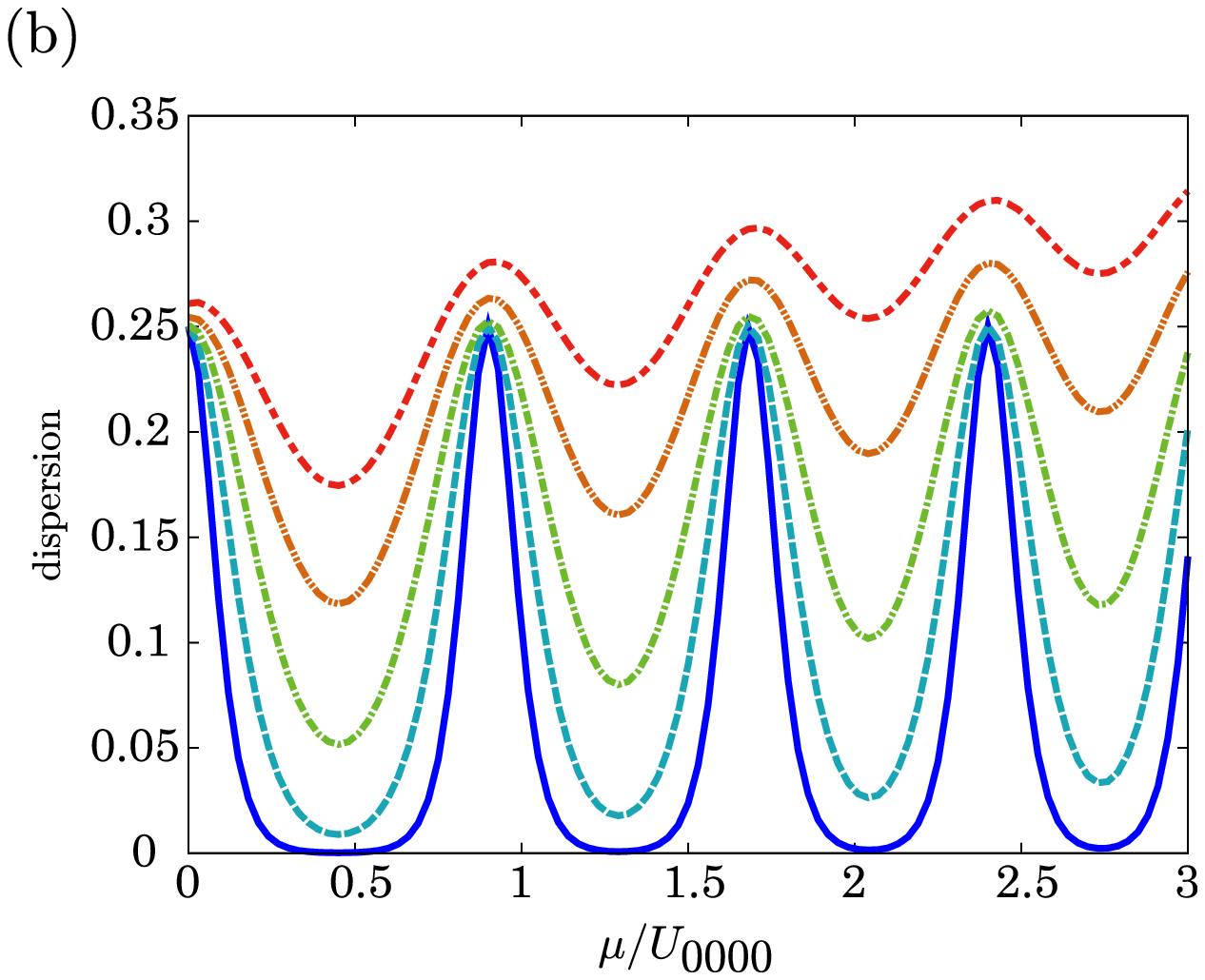}\caption{(Color online) The dispersions of the in-site particle numbers in the zero hopping limits $J=0$ for (a)  the 1D and (b) the P1D models with $\Delta E_y/U_{0000}=3.0\,$.}\label{bunsanT}
\end{figure}

\begin{figure}
\includegraphics[width=4.4cm]{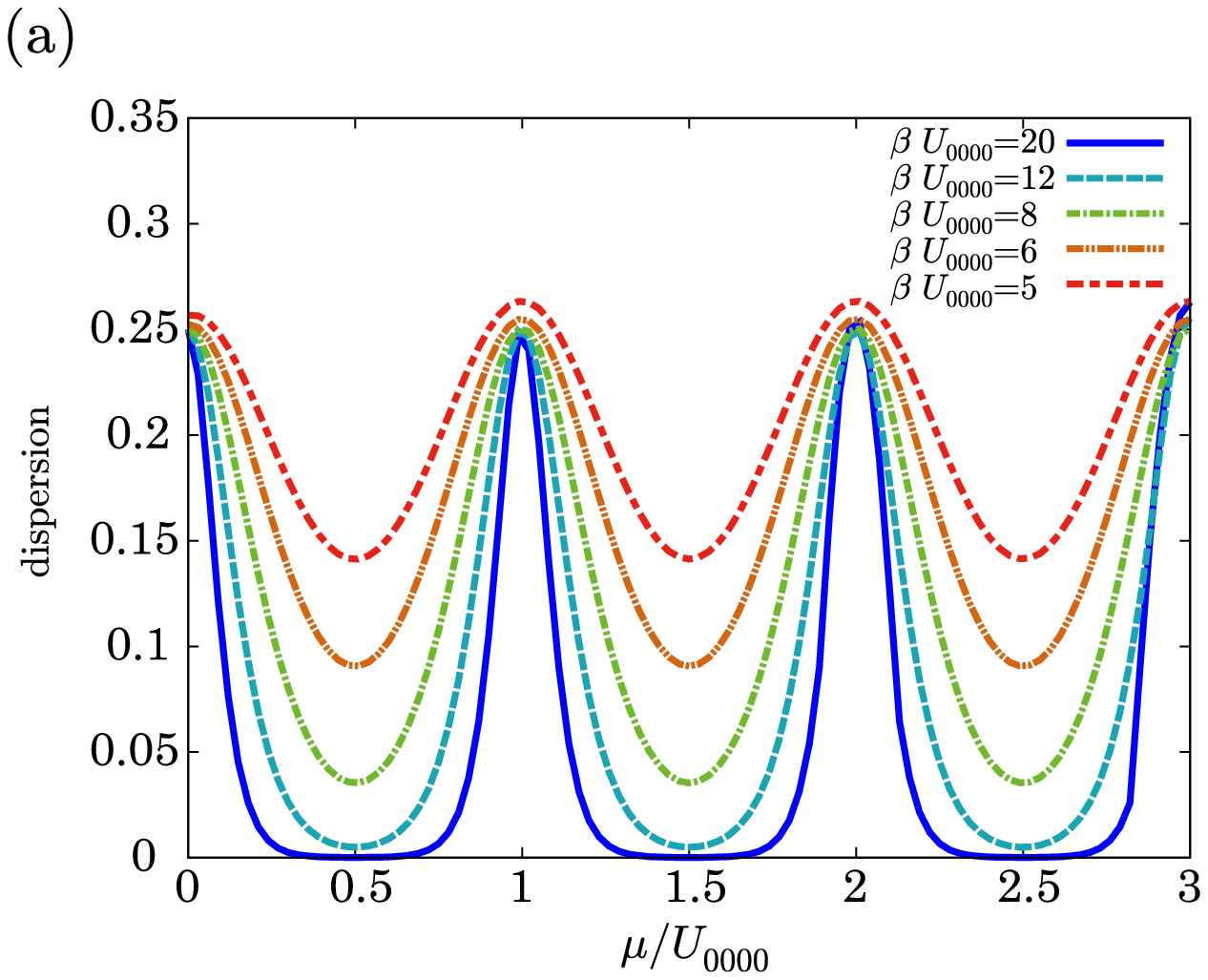}\includegraphics[width=4.4cm]{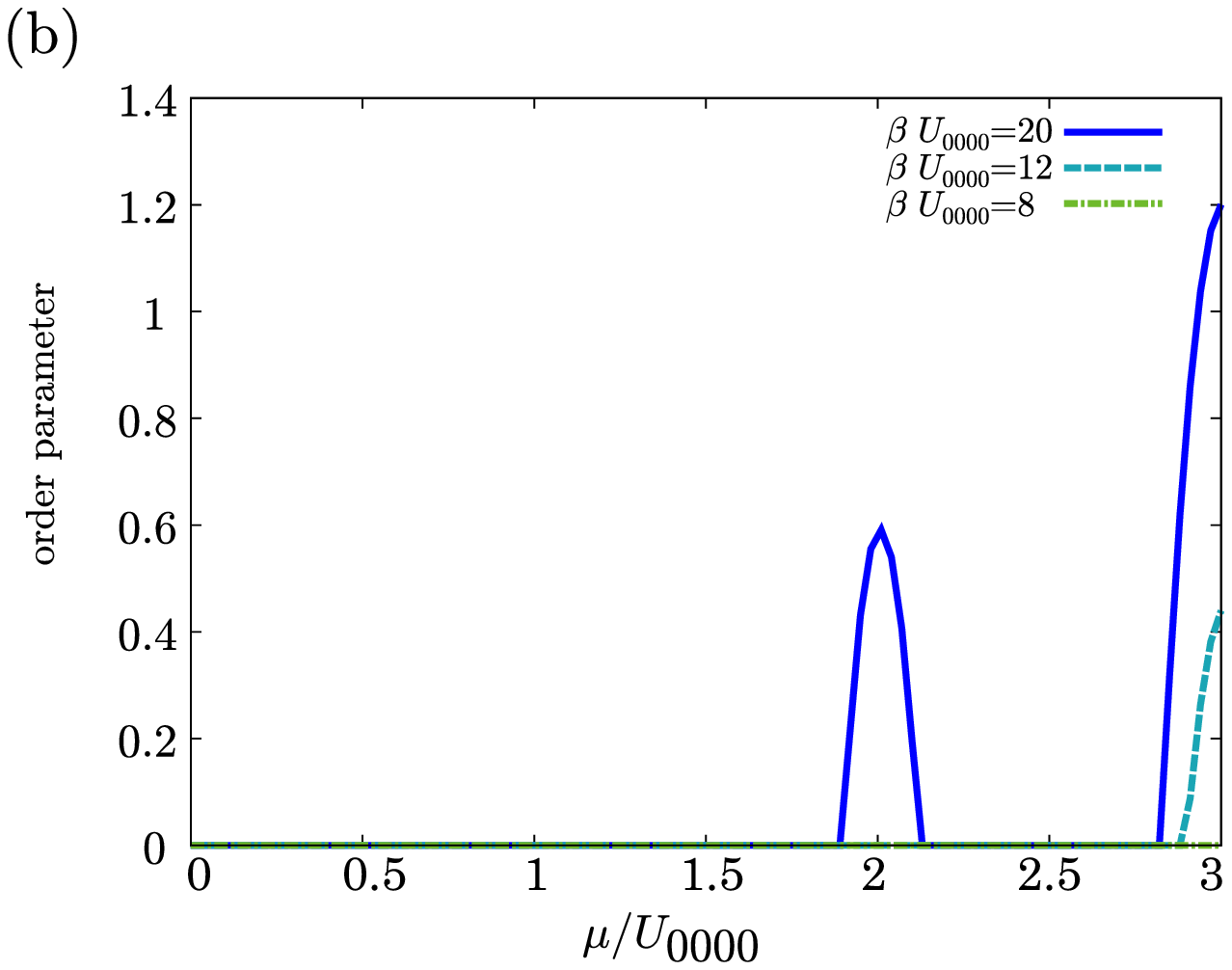}
\includegraphics[width=4.4cm]{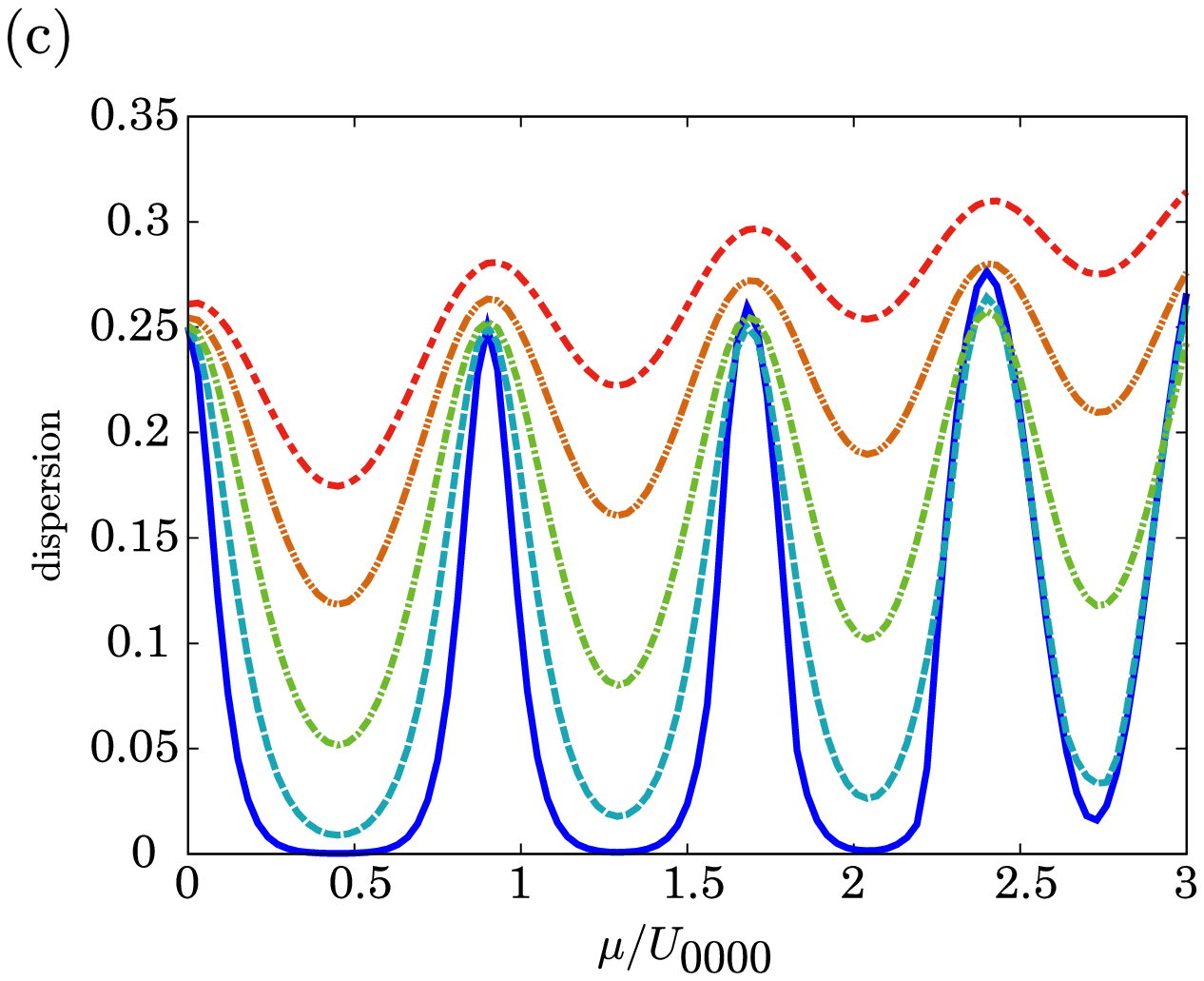}\includegraphics[width=4.4cm]{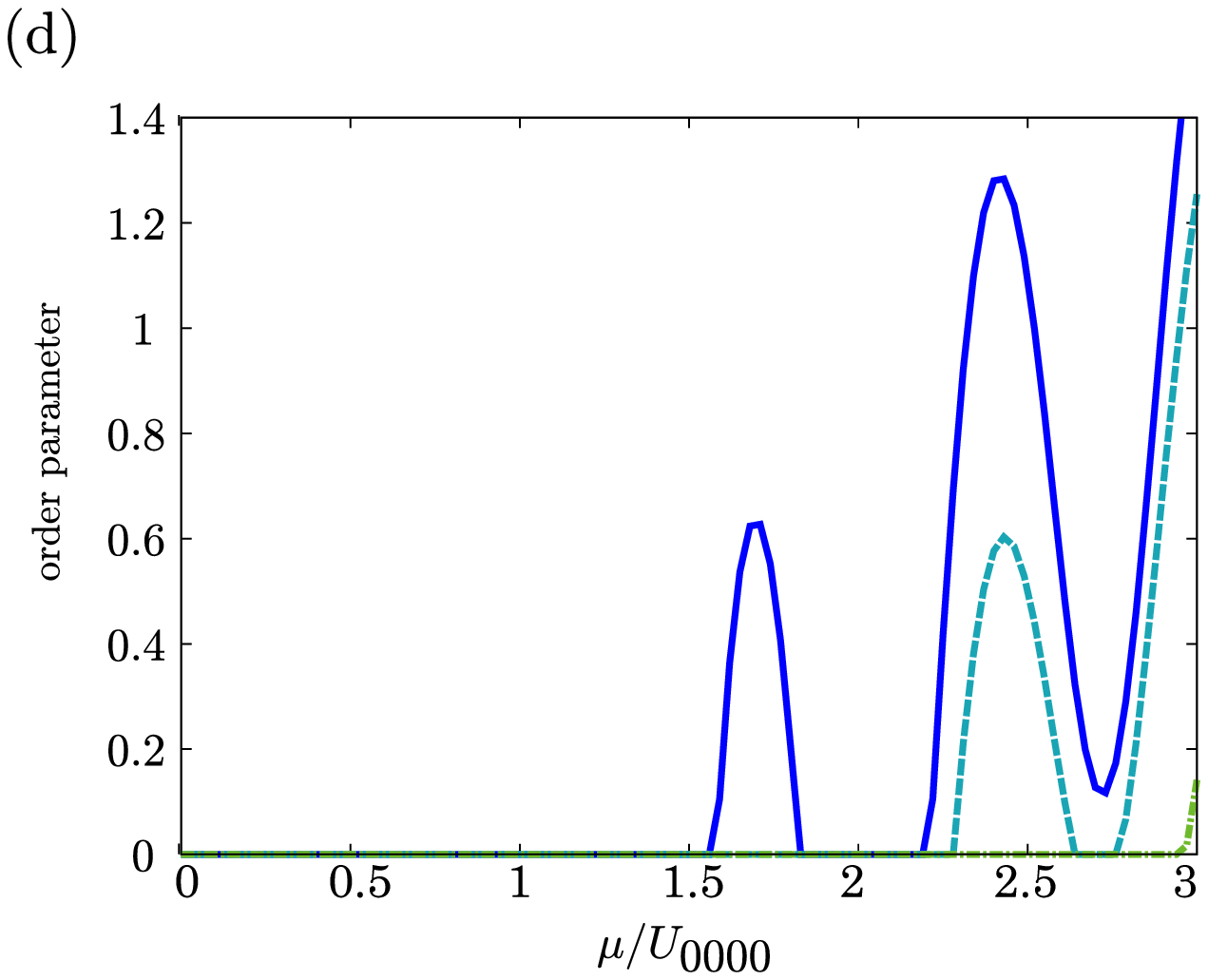}
\caption{(Color online) The dispersions of the in-site particle numbers (a) and the absolute square of the order parameter (b) with $J/U_{0000}=0.02$ for the 1D model. Figures (c) and (d) represent those for the P1D model respectively.}\label{bunsan22T}
\end{figure}

To investigate the temperature dependence of the dispersion in more detail, we show the dispersions for small $J$, namely for $J=0$ in Fig.~\ref{bunsanT} and for $J/U_{0000}=0.02$  in Fig.~\ref{bunsan22T}.
  Comparing the results for the P1D model in Figs.~\ref{bunsanT}-(b) and \ref{bunsan22T}-(c) with those for the 1D model in Figs.~\ref{bunsanT}-(a) and \ref{bunsan22T}-(a), we see that 
 the dispersion for the P1D model
 is larger than that for the 1D model. In addition, the dispersion for the 1D model is independent of the chemical potential $\mu$, but that for the P1D model depends on $\mu$ in such a manner that it is larger at larger $\mu$. The order parameter for the P1D model begins to be non-vanishing in smaller $\mu$ region than for the 1D model, as in Fig.~\ref{bunsanT}-(b) and (d). 
At higher temperature, the order parameter becomes smaller and the region in the normal-liquid phase expands equally for the 1D and P1D models in Fig.~\ref{bunsan22T}.    

\begin{figure}
\includegraphics[width=3.9cm]{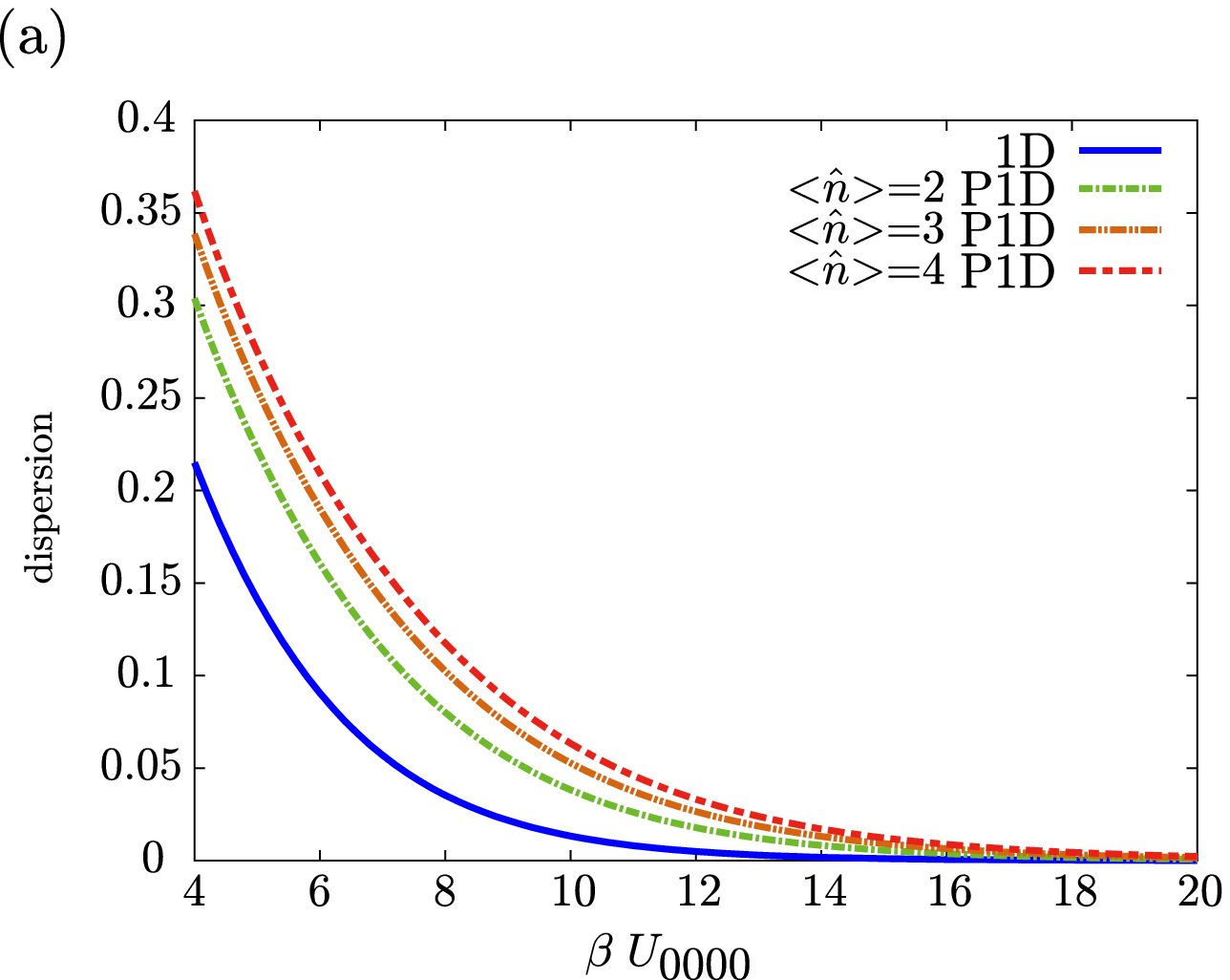}\includegraphics[width=4.3cm]{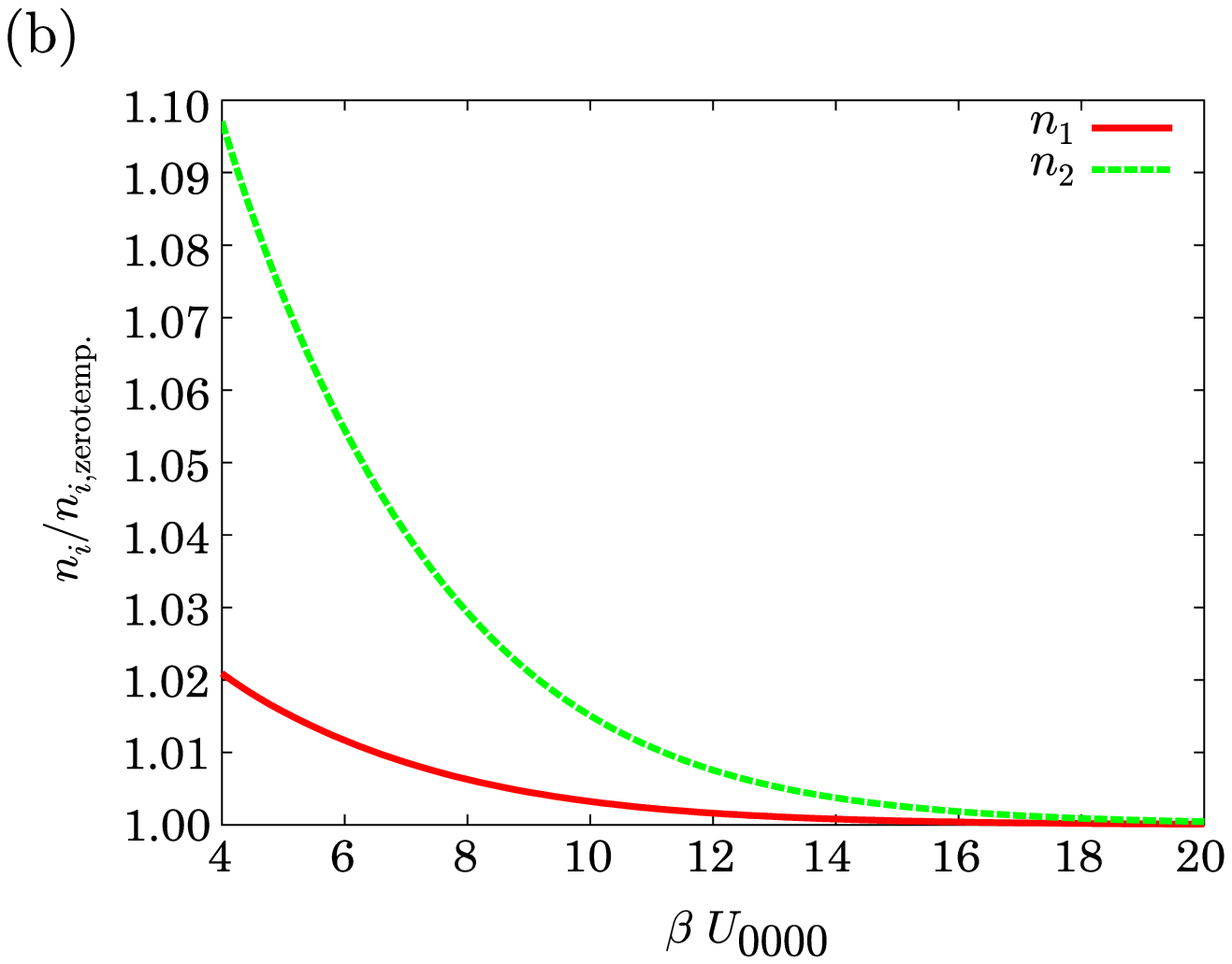}
\caption{(Color online) (a) The dispersions of the average in-site particle number with $\langle\hat{n}\rangle=2, 3, 4$ and $\Delta E_y/U_{0000}=3.0$ for the P1D and 1D models. The dispersion for the 1D model is independent of the average in-site particle number $\langle\hat{n}\rangle$\,.
(b) The ratios of the excited particle numbers at finite temperature $n_i$ to those at zero temperature $n_{i,T=0}$ in the average in-site particle number $\langle\hat{n}\rangle=3$ for the P1D model with $\Delta E_y/U_{0000}=3.0$.}\label{ryuusiT}
\end{figure}

Figure~\ref{ryuusiT}-(a) shows the dispersions of the average in-site particle number with $\langle \hat{n}\rangle=2,3,4$ and $J=0$\,.
In both of the 1D and P1D models,
 the dispersions increase as the temperature rises, 
 but the increasing rate of the dispersion for the P1D model
 depends on the average particle number $\langle \hat{n}\rangle$.
We also illustrate the temperature dependence of the increasing rates of the excited particle numbers for $\ell=1,\, 2$ in Fig.~\ref{ryuusiT}-(b)\,.
 One can find in Fig.~\ref{ryuusiT} that the increase in the excited particle numbers synchronizes with that in the dispersions. This suggests 
that the excited particles contribute to the large dispersions in the P1D model.


\begin{table}
\caption{In this table, $E_m^{\text{(1D)}}$ and $E_{m}^{\text{(P1D)}}$ represent
 the $m$-th eigenvalues of the Hamiltonian for the 1D and P1D
systems with
 $\langle \hat{n}\rangle =3$ and $\beta U_{0000}=10$. These values are respectively the eigenvalues of the grand canonical Hamiltonian $\hat{H}-n\mu$, so the values can be negative.
 The symbol $\Delta E_{m-1,m}$ represents the spacing between energy levels,
 $\Delta E_{m-1,m}=E_m-E_{m-1}$\,.
}\label{tukue-1}
\begin{ruledtabular}
\begin{tabular}{ccccc}
$m$&\hspace{-0.5pt}0&\hspace{-0.5pt}1&\hspace{-0.5pt}2&\hspace{-0.5pt}3 \\
\colrule
$E_m^{\text{(1D)}}/U_{0000}$&
$\hspace{-1.2pt}-4.50$&
$\hspace{-1.2pt}-4.00$&
$\hspace{-1.2pt}-4.00$&
$\hspace{-1.2pt}-2.50$\\

$\Delta E_{m-1,m}^{\text{(1D)}}/U_{0000}$ &&
$\hspace{-0.5pt}0.500$&
$\hspace{-0.5pt}0.00$&
$\hspace{-0.5pt}1.50$\\

$ E_{m}^{\text{(P1D)}}/U_{0000}$&
 $\hspace{-1.2pt}-3.54$&
 $\hspace{-1.2pt}-3.19$&
$\hspace{-1.2pt}-3.19$&
$\hspace{-1.2pt}-2.15$\\

$\Delta E_{m-1,m}^{\text{(P1D)}}/U_{0000}$ &&
 $\hspace{-0.5pt}0.359$&
$\hspace{-0.5pt}0.00$&
$\hspace{-0.5pt}1.04$
\end{tabular}
\end{ruledtabular}
\end{table}
To clarify the rapid increase of the dispersions for the P1D model,
 we investigate the eigenvalues of the Hamiltonian,
\be
\hat{H}^{\text{(M)}}|\Psi_m\rangle=E_m|\Psi_m\rangle\,,
\ee
for the 1D and P1D models.
In Table.\,\ref{tukue-1}, the eigenvalues of the Hamiltonian
 and the energy-level spacings are shown.
One finds that the energy-level spacing for
 the P1D model $\Delta E_{m-1,m}^{\text{(P1D)}}$
 is narrower than that for the 1D model $\Delta E_{m-1,m}^{\text{(1D)}}$. 
The narrow energy-level spacing enhances the contributions of the higher energy eigenstates $m=1, 2, \cdots$ at finite temperature, so that the dispersions become large then.
In addition, it turns out for $\langle \hat{n}\rangle =3$ 
that the eigenstates belonging to the eigenvalues $E_1$ and $E_2$ are those with the in-site particle numbers $n=4$ and $n=2$, respectively, while the eigenstate to belonging to $E_0$ is that with $n=3$\,. The number of the excited particles grows as the in-site particle number does according to Table~\ref{joutai}.  Thus the increase of the excited particles in Fig.~\ref{ryuusiT}-(b) reflects the contribution of the eigenstate with $E_1$, equivalently  of the state of the in-site particle number $n=4$.

The narrow energy-level spacing arises from the fact that the on-site interaction coefficients involving the excited particles are small. This is explained in the following way. As the fraction of the excited particle number becomes large at the large average in-site particle number $\langle \hat{n}\rangle$, the effective repulsive interaction becomes small and so are the eigenvalues $E_m$ $(m=1,2,\cdots)$\,.


The above discussions indicate that the presence of the excited states affects the robustness
 of the Mott phase definitely and essentially.
Although the approximation of the 1D model
 is valid for very low density temperature,
 the effects of the excited states can not be neglected 
for describing more general situations.

\section{Summary and discussion}\label{summary}
In this paper, we analyzed SF-MI transition at zero and finite temperature for the P1D Bose-Hubbard model. 
The P1D model, which differs from the 1D model, contains the transition terms, which are similar to the spin transition and the transition between the Bloch bands.  
 Specifically, the form of the multiband Bose-Hubbard Hamiltonian corresponds to that of the P1D Bose-Hubbard Hamiltonian, aside from the values of the on-site interaction coefficients $U_{\ell_1\ell_2\ell_3\ell_4}$ and the hopping coefficients.  
We expect that our analysis is also useful in analyzing the multiband model.

 In zero temperature analysis, the SF-MI phase diagram for the P1D model is different from that for the 1D model even for the relatively tight confinement.
The major difference is the shrinkage of the Mott lobes in the $\mu$ direction.
The effect is attributed to decreasing the effective on-site interaction by the transition between the ground and excited states. 
The transition produces the fluctuation of the particle number in each $\ell$-state in the Mott lobes. 
The decrease of the on-site interaction enhances the hopping relatively, which leads to the shrinkage of the Mott lobes in the $J$ direction as that in the $\mu$ direction.
We note that the 1D model can not take account of these effects by adjusting the parameters.

At finite temperature, the  dispersion of the in-site particle number increases as the temperature becomes higher. In addition, the region of the normal-liquid phase surrounding the Mott lobes expands. These effects occur both for the 1D and P1D models due to the thermal fluctuation. 
The result for the 1D model corresponds to Ref.~\cite{finitemp}. We note that the increase of the dispersion for the P1D model is enhanced more than that for the 1D model.
The tendency is more noticeable at higher temperature due to the increase of thermal fluctuations.
These results are attributed to  the weaker on-site interaction coefficients involving the excited states.  
 
It have been seen that the robustness of the MI phase is affected at finite temperature by the presence of the excited particles.
 From the view of the local density approximation in Fig.~\ref{2d-8}\,, the P1D Bose-Hubbard model
 predicts that the Mott shell collapses  more quickly than the 1D model does. 
 The validity of the 1D model is restricted to very low density and temperature for realistic cold atomic gases. At high density and/or temperature, one has to 
to consider the excited states coming from the confinement potential.

  \addcontentsline{toc}{section}{\numberline{Reference}}

\end{document}